\documentclass[manuscript,screen,nonacm]{acmart}
\usepackage{listings}
\definecolor{keywordcolor}{RGB}{152, 104, 1}
\definecolor{typecolor}{RGB}{1, 132, 188}
\definecolor{fncolor}{RGB}{80, 161, 79}
\definecolor{typecodecolor}{RGB}{5,56,107}
\newcommand{\code}[1]{\textcolor{typecodecolor}{\texttt{#1}}}
\definecolor{commentcolor}{RGB}{88, 71, 148}
\definecolor{opcolor}{RGB}{228, 86, 73}
\usepackage{listings}
\usepackage[inkscapelatex=false]{svg}
\usepackage{tikz-cd}
\lstdefinestyle{basicstyle}{
    columns=fullflexible,
    basicstyle=\ttfamily,
    breakatwhitespace=false,
    breaklines=true,
    keepspaces=true,
    showspaces=false,
    showstringspaces=false,
    showtabs=false,
    tabsize=2,
    xleftmargin=20pt,
    framexleftmargin=7pt,
}
\lstdefinelanguage{hs}{%
  morekeywords={lala},
  morekeywords=[2]{Map,Vertex,Set,Dist,PriorityQueue,Just,Nothing},   %
  morekeywords=[3]{if,then,else,let,in,case,of,where,data,forall,otherwise,class,instance,newtype,type},
  morekeywords=[4]{dijkstra,lookup,map,put,merge,popQueue},
  otherkeywords={::,+,.,->,@,==,<-,>,:~~:,=,*,[,]},
  morecomment=[l][\color{commentcolor}]--,
  literate={%
    {lambda}{{\color{opcolor}$\lambda$}}{1}
    {->}{{\color{opcolor}$\to$}}{2}
    {<=}{{\color{opcolor}$\leq$}}{1}
  },
  keywordstyle=\color{opcolor},
  keywordstyle=[2]{\bfseries\color{typecolor}}, %
  keywordstyle=[3]{\bfseries\color{keywordcolor}},
  keywordstyle=[4]{\bfseries\color{fncolor}},
  keepspaces
}[strings]%
\lstdefinelanguage{fpop}{%
  morekeywords={lala},
  morekeywords=[2]{Vertex,Dist,Character,State,Partition,Symbol,Pos,Weight},   %
  morekeywords=[3]{relation,rule,order,forall,in},
  morekeywords=[4]{edge,distTo,init,addDist,accepts,notAccepts,isStart,reaches,notAcceptsAll,notRejectsAll,distinguished,hyperEdge,parse,epsilonProduction,lookaheadProduction,andProduction,concatenationProduction},
  otherkeywords={::,+,.,->,@,==,<-,>,:~~:,=,*,[,]},
  morecomment=[l][\color{commentcolor}]--,
  literate={%
    {lambda}{{\color{opcolor}$\lambda$}}{1}
    {->}{{\color{opcolor}$\to$}}{2}
    {<=}{{\color{opcolor}$\leq$}}{1}
    {-->}{{\color{opcolor}$\Rightarrow$}}{2}
    {_}{{\color{commentcolor}$\bullet$}}{1}
  },
  keywordstyle=\color{opcolor},
  keywordstyle=[2]{\bfseries\color{typecolor}}, %
  keywordstyle=[3]{\bfseries\color{keywordcolor}},
  keywordstyle=[4]{\bfseries\color{fncolor}},
  keepspaces
}[strings]%

\AtBeginDocument{%
  }

\received[Updated]{1 June 2025}

\begin{document}

\title{Fixed-Point-Oriented Programming: A Concise and Elegant Paradigm}

\author{Yong Qi Foo}
\email{yongqi@nus.edu.sg}
\orcid{0009-0007-9357-0158}
\author{Brian Sze-Kai Cheong}
\email{brian.cheong@u.nus.edu}
\orcid{0009-0009-2698-1314}
\author{Michael D. Adams}
\authornote{Principal Investigator}
\email{https://michaeldadams.org}
\orcid{0000-0003-3160-6972}
\affiliation{%
  \institution{National University of Singapore}
  \country{Singapore}
}

\authorsaddresses{}

\begin{abstract}
Fixed-Point-Oriented Programming (FPOP) is an emerging paradigm designed to streamline the implementation of problems involving self-referential computations.
These include graph algorithms, static analysis, parsing, and distributed computing---domains that traditionally require complex and tricky-to-implement work-queue algorithms.
Existing programming paradigms lack direct support for these inherently fixed-point computations, leading to inefficient and error-prone implementations.

This white paper explores the potential of the FPOP paradigm, which offers a high-level abstraction that enables concise and expressive problem formulations.
By leveraging structured inference rules and user-directed optimizations, FPOP allows developers to write declarative specifications while the compiler ensures efficient execution.
It not only reduces implementation complexity for programmers but also enhances adaptability, making it easier for programmers to explore alternative solutions and optimizations without modifying the core logic of their program.

We demonstrate how FPOP simplifies algorithm implementation, improves maintainability, and enables rapid prototyping by allowing problems to be clearly and concisely expressed.
For example, the graph distance problem can be expressed in only two executable lines of code with FPOP, while it takes an order of magnitude more code in other paradigms.
By bridging the gap between theoretical fixed-point formulations and practical implementations, we aim to foster further research and adoption of this paradigm.
\end{abstract}

\maketitle

\section{Introduction}

Fixed-Point-Oriented Programming (FPOP) is an emerging programming paradigm that could revolutionize how developers write code for problems involving cycles or other types of self-reference and bring an order of magnitude improvement in programmer productivity.
Examples of such problems include parsing, static analysis, type-checking, graph algorithms, automata minimization, and distributed computing, and we illustrate some of these in Section~\ref{sec:examples}.
However, this paradigm is underexplored, leaving it ripe for rapid advancements and achievements.

As an example of the mismatch between existing paradigms and these types of problems, consider that while array-based problems are naturally solved by iteration and tree-based problems are naturally solved by recursion, what is the natural paradigm for graph-based problems?  The order in which to traverse the vertices of the graph may not be a natural fit for either array iteration or tree recursion.
For example, some vertices might need to be visited multiple times, and the data associated with them (e.g., graph distances) might need to be refined and improved multiple times (e.g., updated with shorter distances) with dynamically determined calculation orders and potentially cyclic dependencies.

It turns out that the natural paradigm for these is the calculation of least fixed points of monotonic transfer functions over lattices.
These are implemented by what are called work-queue or work-list algorithms.
Unfortunately, implementing work-queue algorithms in traditional languages is often tedious and results in complex and brittle programs.
Programmers are forced to write long and complicated programs, only to discover large changes need to be made when the specification evolves or optimizations are desired.
The result of this tedium is a high barrier to implementing, exploring, and optimizing these algorithms.

Given these challenges, an alternative approach is to rethink how we express computation, shifting from imperative work-list management to a more declarative, fixed-point-driven paradigm.
Rather than manually orchestrating iteration order and updates, this perspective treats problem-solving as defining equations over lattices, where solutions emerge naturally through fixed-point computations.

\section{Examples}\label{sec:examples}

To better illustrate the usefulness of FPOP, we offer practical examples for graph distance, automata minimization, and parsing.
We also show the equivalent Haskell implementation of the first example to show how much effort FPOP saves the programmer, but for the sake of brevity, we omit this for the other examples and also elide some less important declarations.\footnote{Readers familiar with Datalog \cite{datalog} might note some similarities to our language.
However, Datalog restricts relations to discrete lattices or power sets of Cartesian products of values.
This limits Datalog's suitability for relations involving non-discrete lattices, such as the distances in the \code{distTo} relation used in the graph distance problem.
We further discuss additional potential enhancements to FPOP that Datalog does not possess in Section \ref{related-work}.}

\subsection{Graph Distance}

This example computes the shortest path between a starting vertex and a destination vertex in a graph with edges of various distances.
Traditionally, a programmer would implement Dijkstra's algorithm \cite{dijkstra}, such as the below written in Haskell.

\begin{lstlisting}[language=hs]
dijkstra :: Map Vertex (Set (Vertex, Dist))
         -> Map Vertex Dist
         -> PriorityQueue (Dist, Vertex)
         -> Map Vertex Dist
dijkstra edges dists work =
    case popQueue work of
        Nothing -> dists -- all work done so we have reached the fixed-point
        Just ((d1, v1), work') ->
            let oldD1 = Map.lookup dists v1 in
            if   oldD1 <= d1
            then dijkstra edges dists work'
            else let dists' = Map.put dists v1 d1
                     addedWork = map (lambda(v2, d2) -> (d1 + d2, v2)) (Map.lookup edges v1)
                     work'' = PriorityQueue.merge work' addedWork
                 in  dijkstra edges dists' work''
\end{lstlisting}

There are several things of note.
First, readers unfamiliar with Dijkstra's algorithm may have difficulty understanding this implementation because the boilerplate involved in the work-queue style of programming dominates the code and obscures both the purpose of the code (computing graph distances) and the key algorithmic optimization (that elements are pulled from the queue based on distance).
Second, this implementation is brittle.
Several design choices were made in this implementation (e.g., the representation of edges as a map between a vertex and a set of vertex-distance pairs), and even minor changes to these choices require significant changes throughout the code.
Third, avenues for further optimization are opaque.
Aside from minor optimizations based on Haskell's semantics, optimizations at a higher, algorithm level may require major changes to the code.
As a result, exploring algorithm-level optimizations is difficult and time-consuming.

Contrast this with implementing Dijkstra's algorithm using FPOP.
The user first declares relations, describing how things relate to each other.
For example, the \code{edge} relation gives the distance between two vertices, and the \code{distTo} relation gives an upper bound on the distance to a vertex from the start vertex.
These are declared with the following lines.\footnote{In a practical language, terms such as \code{relation} and \code{order} can be condensed into \code{rel} and \code{ord}.
However, here, we leave it as \code{relation} for clarity.}

\begin{lstlisting}[language=fpop]
-- Note that these relations use mixfix notation
relation edge _ _ = _: Vertex, Vertex, Dist -- distance of an edge
relation distTo _ <= _: Vertex, Dist -- upper bound on distance to a vertex
\end{lstlisting}

Next, the user defines rules for inferring new instances of those relations (i.e., facts).
For example, the distance from the starting node to itself is zero (the \code{init} inference rule), and an edge going to a vertex places an upper bound on the distance to that vertex (the \code{addDist} inference rule).
Again, these can be concisely given with the following lines.
\begin{lstlisting}[language=fpop]
rule init: distTo start <= 0
rule addDist: distTo v1 <= d1, edge v1 v2 = d2, d1 + d2 <= d --> distTo v2 <= d
\end{lstlisting}
With just these two executable lines of code, we have defined graph distance, and this is sufficient for the language to produce a runnable implementation.
To get the asymptotic performance of Dijkstra's algorithm, we just have to ensure that when multiple instances of the \code{addDist} rule can be applied, the program always applies the one involving the shorter distance first.
Fortunately, this optimization can be implemented by the compiler in a relatively straightforward analysis based on dependencies between rules and which rule applications can make other applications obsolete (e.g., tighter bounds obsolete looser bounds).
However, even in the absence of such an automatic optimization, users can easily provide their own directives specifying this as follows.

\begin{lstlisting}[language=fpop]
order: d11 <= d12 --> addDist { d1 = d11 } <= addDist { d1 = d12 }
\end{lstlisting}

And that is all! With just a few lines of code, Dijkstra's algorithm is fully implemented!

Also, because the program is so short, amending its design or expanding the problem domain is straightforward.
While one could use an existing implementation of Dijkstra's algorithm from a library, expressing it so easily opens opportunities for modification.
For example, in road mapping, one may want alternate routes but be unsure how best to produce alternate paths that are different enough from the primary path without being too much longer.
Expressing the problem so simply makes it easy to try modifications and make adjustments until one finds a method that produces good alternate routes.

\subsection{Finite-State Automata Minimization}

Automata minimization involves several subproblems, including the following examples of what states are reachable, what states reject or accept all strings, and what states are distinguishable.

We define our automata in terms of the following two relations:

\begin{lstlisting}[language=fpop]
relation edge: Character, State, State -- the character goes from state to state
relation accepts: State -- the state is an accepting state
\end{lstlisting}
Reachability is then defined with the following two rules.

\begin{lstlisting}[language=fpop]
rule: isStart s --> reaches s -- the start state is reachable
rule: reaches s1, edge c s1 s2 --> reaches s2 -- states from reachable states are reachable
\end{lstlisting}

Which states reject all strings (i.e., are ``dead'' states) can be similarly defined as follows.
Note that since this property is co-inductive instead of inductive (i.e., it is an ``optimistic'' property), we define its negation, \code{notRejectsAll}.
The definition for \code{notAcceptsAll} is similar.
\begin{lstlisting}[language=fpop]
rule: accepts s --> notRejectsAll s -- an accepting state does not reject all
rule: edge c s1 s2, notRejectsAll s2
      --> notAcceptsAll s1 -- a state (s1) going to a state (s2) that does not reject all,
                         -- does not itself reject all
\end{lstlisting}
Finally, we can determine which states are distinguishable using essentially Hopcroft's algorithm \cite{10.5555/891883}.
In other words, accepting and rejecting states are distinguishable, and states that go with the same character to distinguishable states are distinguishable:
\begin{lstlisting}[language=fpop]
relation distinguished: Partition(State)
rule: accepts s, notAccepts s' --> distinguished s s'
rule: edge c s1 s2, edge c s1' s2', distinguished s2 s2' --> distinguished s1 s1'
\end{lstlisting}
Note that while the other relations in this example are defined over a power-set lattice (i.e., for what arguments a relation is true), \code{distinguished} is defined over a partition lattice.
This represents the partition in linear space instead of the quadratic space taken by a set of pairs of distinguishable states, as one would do in Datalog.

Even generalizing from string automata to tree automata is easy, even though tree automata are usually more complicated to implement.
For example, \code{notRejectsAll} swaps out the \code{edge} relation from a string automaton with \code{hyperEdge} representing a production in a tree automaton, as in the following.
\begin{lstlisting}[language=fpop]
rule: hyperEdge s1 c ss, forall s in ss. notRejectsAll s --> notRejectsAll s1
\end{lstlisting}

\subsection{Parsing}

Many variants of the parsing problem are easily defined in this language.
Context-free parsing is defined with the following relation and rules for a grammar in Chomsky normal form \cite{chomsky}.

\begin{lstlisting}[language=fpop]
relation parse: Symbol, Pos, Pos -- the symbol parses the input between the positions
rule: concatenationProduction nt a b, parse a i j, parse b j k --> parse nt i k
rule: epsilonProduction nt --> parse nt i i
\end{lstlisting}

More complex operators such as \emph{lookahead} can easily be added:
\begin{lstlisting}[language=fpop]
rule: lookaheadProduction nt a, parse a i j --> parse nt i i
\end{lstlisting}
Even operators considered exotic are no trouble, such as Boolean grammar intersection \cite{booleangrammars}, where a symbol parses only if two other symbols parse the same region of a string:
\begin{lstlisting}[language=fpop]
rule: andProduction nt a b, parse a i j, parse b i j --> parse nt i j
\end{lstlisting}
Adding a weight to compute the cost of the ``best'' parse out of multiple alternatives requires only adding a parameter to the parse relation:
\begin{lstlisting}[language=fpop]
relation parse _ _ _ <= _ : Symbol, Pos, Pos, Weight
rule: concatenationProduction nt a b w, parse a i j <= w1, parse b j k <= w2
      --> parse nt i k <= (w + w1 + w2)
rule: epsilonProduction nt w --> parse nt i i <= w
\end{lstlisting}

In addition, different parse algorithms (Early, CYK, LR, etc.) frequently amount to different scheduling strategies and dependency analyses for when to apply different inference rules.
Also, many analyses of grammars (e.g., emptiness and nullability) are easily expressed this way.
Even LR parser generation can be expressed in only a few inference rules (i.e., state exploration and item-set closure).

\subsection{Discussion}

When written in fixed-point form, these example problems seem almost trivial, but from experience, each one of these problems, when written by hand, is significantly more complicated to implement, debug, optimize, maintain, and have confidence in its correctness.
The almost-trivial ease with which these examples are expressed in FPOP highlights the gains it can bring to expressiveness and flexibility.

In addition to the examples given here, fixed-point-oriented problems occur in many domains of interest to academia and industry.
For example:

\begin{itemize}
    \item \textbf{Static analysis}, used in formal verification and proof systems, relies on the identification of fixed points to determine the behavior of programs and is widely used to verify the correctness, safety, and security of software.
    \item \textbf{Parallel and distributed computing} can take advantage of how fixed-point computations can be split, computed in parallel, and then joined.
    For example, conflict-free replicated data types (CRDTs) operate on lattices, with updates being joins on those lattices.
    \item \textbf{Rendering and layout algorithms} such as Knuth-Plass line breaking \cite{https://doi.org/10.1002/spe.4380111102} are well suited to being described in terms of fixed points.
    \item \textbf{Graph algorithms} frequently use fixed points.
    In addition to the already mentioned Dijkstra's algorithm, examples include strongly connected components \cite{robert1972}, graph dominators~\cite{10.1145/357062.357071}, and finding maximum flows \cite{ford2}.
\end{itemize}

FPOP expresses complex problems concisely, avoids the repetitive boilerplate code of traditional work queues (which in turn reduces the likelihood of bugs), and simplifies problem definition by focusing on high-level abstractions rather than low-level implementation details.
This reduces implementation time and frees programmers to more easily explore the design space of their code.

These problems benefit from a high-level language for concisely and easily expressing fixed-point problems that generates programs from users' specifications and exposes avenues for optimization.
As such, this paradigm fills in a gap in programming language theory and enables the field to grow more rapidly by accelerating programmer productivity and researcher exploration.

\subsection{Future Directions}

Though some work exists on FPOP \cite{datafun, flix}, it is an underexplored paradigm that our group believes is a fertile area for research.
To that end, our group has begun exploration in this paradigm of fixed-point computation and its supporting theories \cite{fpoponline}.
We aim to develop a fixed-point-oriented language, investigate how a programming language centered around fixed-point principles could potentially be structured and implemented, and discover what natural patterns would arise through programming in a fixed-point-oriented manner.

Notably, fixed-point algorithms typically have flexibility in their calculation order and multiple ways of representing intermediate computations.
Thus, our language will let programmers write high-level code that performs well through user-guided optimizations such as solving order, indexing, and fact representation.
Optimization directives will allow programmers to modify the solution strategy and problem representation without having to modify the inference rules.
This allows programmers to rapidly prototype and test inference rules, then later optimize their code without fear of losing correctness since the inference rules remain largely untouched.
This mitigates the problem of traditional work-queue code often requiring extensive modifications to achieve optimized performance.

In addition, the high level of abstraction at which FPOP code is written encodes the intent of the programmer more directly than traditional work queue algorithms,
which presents an opportunity for the compiler to apply algorithm-level optimizations.

Finally, we aim for FPOP's adoption by the general programming community to solve fixed-point-oriented problems, similar to how parser generators, SMT solvers, and SQL are called from general-purpose languages for domain-specific tasks.
This allows us to develop features and syntax tailored towards fixed-point computations while only requiring the parts of the applications suited for fixed points to be written in the fixed-point language.

\section{Background Theory}
In this section, we provide background information for readers to understand our characterization of the earlier problems as fixed points and to understand our approach and its feasibility.
For brevity, we gloss over some technicalities.

\subsection{Definitions}
\subsubsection{Partial Orders and Lattices} \label{def:lattices}

A \emph{partial order} ($\sqsubseteq$) on a set $L$ is a relation that is reflexive ($x\sqsubseteq x$), transitive ($x\sqsubseteq y$ and $y \sqsubseteq z$ implies $x\sqsubseteq z$), and antisymmetric ($x\sqsubseteq y$ and $y\sqsubseteq x$ implies $x=y$).
The word \emph{partial} indicates that not every pair of elements is necessarily ordered.
A \emph{partially ordered set}, or \emph{poset}, is then the pair $(L, \sqsubseteq)$.
Examples include power sets, numbers (e.g., for graph distance), class hierarchies, and subtyping.

Within a poset, the \emph{join} operation ($x\sqcup y$) gives the smallest $z$ such that $x \sqsubseteq z$ and $y \sqsubseteq z$ (i.e., the least upper bound of $x$ and $y$).
The \emph{meet} operation ($x\sqcap y$) gives the largest $z$ such that $z \sqsubseteq x$ and $z \sqsubseteq y$ (i.e., the greatest lower bound of $x$ and $y$).
A poset is a \emph{lattice} if every pair of elements has both a join and a meet.
Additionally, if a poset has a single \emph{greatest element}, we call that its \emph{top} ($\top$), and if it has a single \emph{least element}, we call that its \emph{bottom} ($\bot$).

\subsubsection{Monotonicity and Fixed Points}
A function $f:L\to M$ between posets $(L,\sqsubseteq_L)$ and $(M,\sqsubseteq_M)$ is \emph{monotonic} if it is order preserving ($x \sqsubseteq_L y \Rightarrow f(x) \sqsubseteq_M f(y)$).
A fixed point of a function $f$ is an $x$ such that $f(x)=x$.
In other words, some $x$ such that applying $f$ to $x$ yields $x$ itself.

\subsubsection{Kleene's Fixed-Point Theorem}
Kleene's fixed-point theorem states that if $(L,\sqsubseteq)$ is a poset and $f$ is monotonic, then under appropriate conditions, one can compute the least fixed point (lfp) of $f$ by repeatedly iterating $f$ over $\bot$ until a fixed point is found.
Greater elements in the lattice are thought of as subsuming and obsoleting lesser elements, and Kleene iteration ``climbs'' the lattice until a fixed point is found.
Everything can also be dualized where the greatest fixed point is sought, lesser elements obsolete greater elements, and iteration ``descends'' the lattice starting from $\top$ to find a fixed point.
Different research areas and communities adopt different conventions in this regard.

\subsection{Applicability of Theory to Practice}
Kleene's fixed-point theorem shows that iteratively inferring new relation instances from known relation instances leads to finding a fixed point.
In our application, the domain $L$ represents the lattice over which relations operate, $x$ represents relation instances that hold, the partial order $\sqsubseteq$ represents when one relation instance subsumes or obsoletes another, and $f$ computes the consequences of the inference rules based on the existing relations.
Combined with optimizations to apply only rules affected by newly learned relation instances, this leads to the classic work-queue algorithm.

Framing problems in this way is not just a theoretical contrivance.
Many algorithms in these domains are already framed in terms of fixed-point computations.
To illustrate how Dijkstra's algorithm can be seen as a fixed-point problem, consider a directed graph $G = (V,E)$, where $A$ is the start vertex.

In order to represent distances, we define a partially ordered set $D = \mathbb{N} \cup \{\infty\}$, where numerically smaller (and thus better) distances are greater in the partial order and $\infty$ represents there being no path.
Then, to represent the distances to all vertices, we define a partially ordered set $\mathcal{D} = V \to D$ that maps each vertex to the shortest known distance from A to that vertex.
The partial order $\sqsubseteq$ on $\mathcal{D}$ is such that for any $d_1, d_2 \in \mathcal{D}$, $d_1 \sqsubseteq d_2$ if and only if for all $v \in V$,
$d_1(v) \sqsubseteq d_2(v)$.
That is, $d_1 \sqsubseteq d_2$ whenever $d_2$ provides a more refined (i.e., numerically less than or equal) estimate of the shortest distances.
The least element ($\bot$) of $\mathcal{D}$ maps all vertices to $\infty$.
This represents where no shortest paths are known.

Given this structure, Dijkstra's algorithm can be interpreted as an iterative process for computing the least fixed point of a monotonic function $f: \mathcal{D} \to \mathcal{D}$ that updates distance estimates according to the shortest paths discovered so far, where $f$ is
\[
f(d)(v) = d(v) \sqcup \bigsqcup_{(u,v)\in E} (d(u) + w(u,v))
\]
where $E$ is the set of all edges and $w(u,v)$ is the distance of the edge $(u,v)$.
This function refines the current estimate by ensuring that each distance satisfies the triangle inequality with respect to its neighboring nodes.

Since $f$ is monotonic with respect to $\sqsubseteq$, the sequence $\bot, f(\bot), f^2(\bot), \cdots$ forms an ascending chain in $\mathcal{D}$.
By Kleene's fixed-point theorem, this sequence converges to the least fixed point $d^*$, which represents the shortest path distances from $s$ to all other nodes.
The function $f$ iteratively refines the distance estimates, effectively ``climbing" the lattice until it stabilizes at the least fixed point, which at that point would contain the shortest distances to all nodes.

Apart from Dijkstra's algorithm, there are many other examples, such as parsing, static analysis, and other graph algorithms.

\paragraph{Parsing}
Parsing context-free grammars is naturally a fixed-point computation based on inference rules.
Any time a symbol parses a region of a string, any production that uses that symbol might be able to parse from it.
For instance, Earley's parser \cite{10.1145/362007.362035} iterates through parse states, adding new states as it goes until no new states are introduced, at which point parsing reaches a fixed point, the parsing is complete, and all possible derivations have been considered.
Similarly, the CYK (Cocke-Younger-Kasami) algorithm \cite{syntaxinuniversaltranslation,cocke,younger,kasami} constructs a parse table by systematically combining smaller substrings into larger ones according to grammar rules.
The table-building process relies on previously filled entries, and the algorithm iterates until the table reaches a stable state---a fixed point where no further updates are necessary.
In attribute grammars, attribute evaluation involves computing attributes for grammar symbols based on semantic rules, often through iterative processes.
These computations converge to a fixed point when all attribute values stabilize, ensuring that all dependencies are resolved and no further changes are needed.
Even the generation of parsers, such as in LR parsing, is based on inferring new states from known states until the states of the push-down machine are fully elaborated.

\paragraph{Static Analysis}
Static analysis tasks, such as type inference, control-flow analysis, and data-flow analysis, are frequently framed in terms of fixed points.
For instance, in data-flow analysis, the objective is to determine how values propagate through a program.
This starts with initial assumptions (e.g., empty sets or default values) and iteratively applies transfer functions across the control flow graph until the results stabilize and no longer change between iterations (i.e., we have reached a fixed point).
This stable state represents the final data-flow information.
Similarly, type inference often involves deducing types through iterative constraint solving that converges to a fixed point where type assignments stabilize and do not change further.
Abstract interpretation uses fixed-point computations by approximating program semantics over an abstract domain.
Here, iterative application of abstract operations continues until a stable approximation---a fixed point---is achieved.
Verifying program properties often involves iterating to compute invariants or properties until they stabilize, ensuring that certain conditions hold across all possible executions.

\paragraph{Graph Algorithms}
Fixed points are widely used in graph algorithms.
In addition to the already mentioned Dijkstra's algorithm, the Bellman-Ford algorithm \cite{bellman,ford1} for finding shortest paths involves iteratively relaxing edges to update path estimates.
This continues until no further improvements can be made.
The algorithms for strongly connected components \cite{robert1972}, graph dominators~\cite{10.1145/357062.357071}, and finding maximum flows \cite{ford2} also involve iterative methods that converge to fixed points.

\section{Approach}
\subsection{Language Design}
At a high level, an FPOP language can be structured around types, lattices, relations, rules, and optimization directives.
We describe each of these in turn, drawing from the examples shown earlier.

\begin{itemize}

\item \textbf{Types:}
Types define the values that fixed-point computations operate over and are used to define lattices.
These can be natively defined within the language, but because our language is a solver, we plan to allow programmers to refer to “foreign” types that are defined in the language calling our solver.
This eases the marshaling of data between the calling program and the solver.

\item \textbf{Lattices:}
Lattices add orders to types and allow the arguments to different instances of a relation to be combined.
For example, in graph distance, we keep only the shortest distance, so we combine multiple distances to the same vertex by keeping only the shortest distance.
Lattices can be defined by directly defining the ordering, join, and meet operators or by several operators, such as ``reversing'' a lattice or taking the power set of a type.
Note that a type can have multiple lattices (e.g., numbers have multiple orders that can be defined).

\item \textbf{Relations:}
Relations define relationships between different values.
These enable the expression of complex dependencies and constraints that formulate and solve fixed-point problems.

\item \textbf{Rules:}
Inference rules are the core construct that drives the computation forward.
Rules infer new relation instances from known relation instances and thus define how new information is inferred from existing information.

\item \textbf{Queries:}
Queries are interfaces that external code can use to ask about the relation instances inferred by the solver.

\item \textbf{Optimization Directives:}
The language exposes optimization directives that provide guidance to various aspects of the solver (e.g., in Dijkstra's algorithm, the order in which to apply instances of the \code{distTo} rule).
Ideally, the compiler can analyze the relations and rules to automatically determine the optimal rule order and relation representation, but for early prototypes of the language or in cases that are particularly complicated, this allows the user the control needed for algorithmic efficiency.

\end{itemize}

\subsection{Architecture}
The compiler architecture can be viewed from several perspectives.
First is the compilation pipeline and how one would integrate it into a calling application.
Second is the API that the calling application would use to interface with the solver.
Third is the internal algorithm that the solver uses.

\subsubsection{Compilation Pipeline}

\begin{figure}[ht]
    \centering
    \centerline{\includesvg[width=0.8\columnwidth]{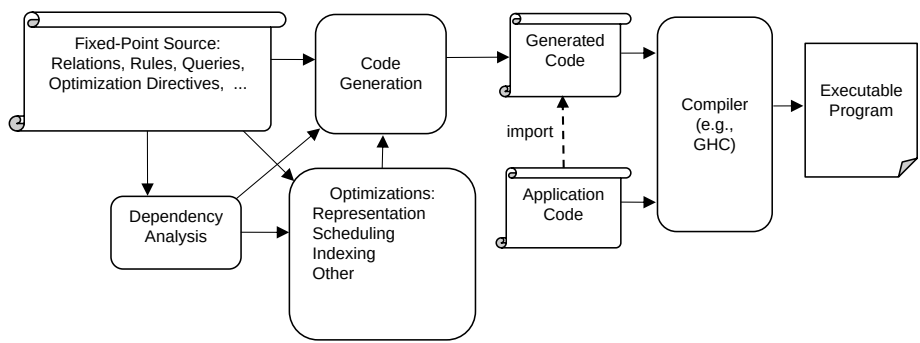}}
    \caption{The compilation pipeline}
    \label{fig:compiliation-pipeline}
\end{figure}

As shown in Figure~\ref{fig:compiliation-pipeline}, the compilation process starts with fixed-point source code written by the user, containing lattices, relations, rules, and so on.
A dependency analysis determines the dependencies between rules.
This is used by the downstream optimizations, which include determining relation representations, ordering which rules should be applied before other rules (scheduling), selecting intermediate caches to more efficiently look up relation instances (indexing), and many other optimizations.
Once all the optimizations have run, the code generator emits optimized Haskell code that implements the problem specified in the fixed-point source code.
This code can then be imported by application code and both be compiled into the final executable program.

We also intend to develop a Template Haskell \cite{10.1145/581690.581691} wrapper that simplifies this process, so users can write fixed-point source code directly within a Template Haskell quasi-quote in their application code.
This makes the build process more lightweight for developers.

\subsubsection{Usage by Calling Code}

\begin{figure}[ht]
    \centering
    \centerline{\includesvg[width=0.8\columnwidth]{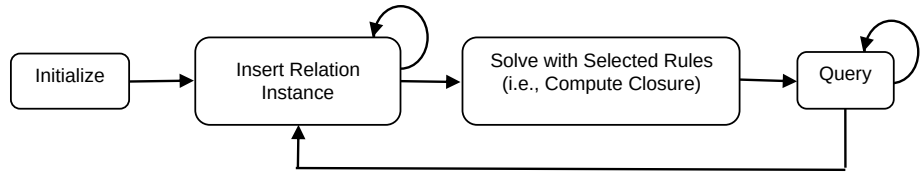}}
    \caption{External interface to the solver}
    \label{fig:usage}
\end{figure}

The code generated by the code generator contains a module exporting a solver over the relations and rules declared in the fixed-point source code.
Application code uses this solver using four types of API functions: initialization, insertion of relation instances, solving, and querying.
This is illustrated in Figure~\ref{fig:usage}.

After initialization, inserting relation instances is how a programmer specifies a particular problem instance to be solved.
For example, with graph distance, one adds instances for the \code{edge} relation to specify the graph to be solved.
Once all appropriate instances are added, the solving function is called, which kicks off the work-queue-based solving algorithm.
Once the work-queue algorithm reaches a fixed point, the user can then query the resulting relation instances that were discovered.

Because of the lattice structure, one can also repeat this process by adding new relation instances even after the solver has found a fixed point and then invoke the solver again.
These subsequent runs of the solver automatically reuse results from previous runs to efficiently compute the new solution.
For example, after computing graph distances and inspecting the result, one could re-run the solver with new edges added, and only the distances affected by those new edges would require re-computation.

\subsubsection{Internal Solver Process}

\begin{figure}[ht]
    \centering
    \centerline{\includesvg[width=0.5\columnwidth]{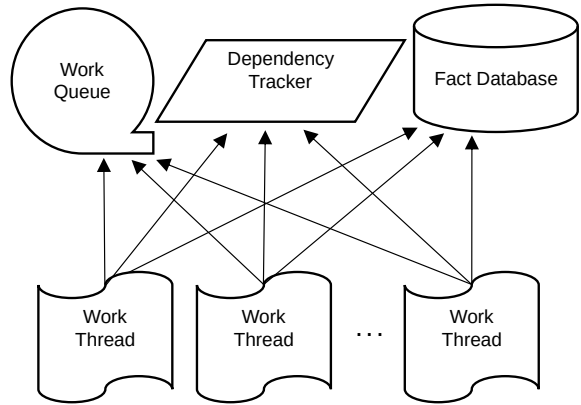}}
    \caption{Internal process when the solver is run}
    \label{fig:internal-solver}
\end{figure}

Within the solver, multiple work threads run in parallel, each of which runs a main loop based on the classic work-queue/work-list algorithm.
They apply rules based on known relation instances to infer new relation instances, and they interact with three primary types of shared data: relation instances, the work queue, and dependency information.
These are shown in Figure~\ref{fig:internal-solver}.

The facts database tracks relation instances inserted by the calling application and what is inferred by the solver.
Facts have a lattice structure based on the lattices used by relations, so we can optimize how they are stored.
For example, if $x\sqsubseteq y$, then the relation instance for $y$ subsumes that of $x$, and $x$ does not need to be stored (e.g., in graph distance, shorter path distances subsume longer path distances).
As another example, if the lattice is over a partition (e.g., in automata minimization), then there are more efficient ways to store things than pairs of dis-equalities.
At the same time, these facts must be stored in a way that allows efficient lookup when applying a rule (e.g., when the \code{addDist} rule uses \code{distTo} and \code{edge} as premises).
This can be accomplished by storing indexes that are updated as new instances are learned, but at the same time, there is a trade-off between the cost to update these indexes and their speedups to lookup.
There are many optimization opportunities in this area, so we anticipate this to be a fertile area of research.

The work queue is a priority queue that tracks new relation instances to be added to the database but whose consequences have not been applied in terms of rules that could be applied as a result of them being added to the database.
They represent work in the form of rules that need to be further applied, and the main loop of the solver amounts to taking tasks from this work queue and applying the rules that are consequences of those tasks.
When these rules are applied, they create more relation instances that go back into the work queue until no new instances are generated by rules and the work queue empties.
At that point, the solver has found a fixed point.
The order in which to pull elements from this queue can affect performance.
For example, Dijkstra's algorithm relies on shorter distances in the queue being processed before longer distances so that the number of times each vertex is processed is minimized.
This is another area where we foresee fertile areas for research and opportunities for optimization, both from optimization directives and automated program analyses.

The dependency tracker improves performance by tracking the dynamic dependencies between rules and relations.
This allows the solver to quickly identify what rules to apply when and how updating certain relations will affect other relations.

Each work thread repeatedly pulls tasks from the work queue in the form of relation instances, adds this relation to the database, and checks whether doing so changes the database.
For example, due to the lattice structure, adding a longer distance to a vertex in the graph distance problem does not change the database.
If it does change the database, then it computes what rules apply because of that update.
The resulting relation instances are then put in the queue to be subsequently processed.
Because of the lattice structures of the fact database, multiple threads can operate simultaneously, which provides an exceptionally easy programming model for parallelism.

\section{Related Work} \label{related-work}

We have explained the advantages of FPOP relative to traditional iteration and recursion, but we also wish to briefly discuss the current landscape for the paradigm of fixed-point-oriented programming languages, including Datalog~\cite{datalog}, Souffl{\'e}~\cite{DBLP:conf/cav/JordanSS16,DBLP:conf/cc/ScholzJSW16}, constraint logic programming (CLP)~\cite{DBLP:journals/jlp/JaffarM94}, Flix~\cite{flix}, and Datafun~\cite{datafun}.

Datalog is a well-established language that supports bottom-up inference of relations, treating relations as sets of tuples.
However, its set-based nature means that multiple instances of a relation have no inherent connection to each other.
For example, Datalog does not take advantage of the fact that a tighter bound on a distance subsumes a looser bound, which limits its applicability to many fixed-point problems.
Constraint Datalog extends the traditional Datalog model by incorporating constraints, which allows for more expressive problem specifications, but it still does not fundamentally address algorithmic optimizations for fixed-point problems.

Flix and Datafun are research languages that extend Datalog's capabilities by supporting lattice-structured relations.
These languages incorporate elements of functional programming and type systems to facilitate reasoning about recursive and fixed-point computations.
Rather than serving as isolated solvers, they provide general-purpose programming capabilities within the fixed-point paradigm.

Our proposed approach aligns with this broader research direction by focusing on FPOP as a dedicated solver within this paradigm.
By taking this approach, we can tailor our language to more directly serve FPOP-specific requirements while allowing developers to integrate our solver into their existing workflows without requiring them to port their entire application to our language.
Additionally, while these languages have explored early work on FPOP, there is still significant design space left to be explored.
Our approach contributes to this space by emphasizing algorithm-level optimizations, providing users with the flexibility to experiment with different implementations through optimization directives without modifying the problem specification.
Our group also aims to extend the types of lattices and relations that can be expressed.
For example, Flix only allows for one lattice argument in a relation, while we allow an arbitrary number of lattice arguments.

\section{Conclusion}

FPOP represents an exciting step forward in programming paradigms, providing a structured and expressive framework for tackling inherently recursive and self-referential problems.
By abstracting fixed-point computations into high-level inference rules and allowing for algorithm-level optimizations, FPOP empowers developers to write clear, maintainable, and performant code.

The challenges associated with implementing work-queue algorithms manually, such as verbosity, brittleness, and difficulty in optimization, are greatly alleviated through FPOP's declarative approach.
This not only benefits research in programming languages but also has practical implications for a wide range of domains, including formal verification, graph processing, parsing, and distributed computing.

We encourage further research and development in the space of fixed-point-oriented programming, exploring additional optimizations, expanding upon the expressiveness of existing fixed-point-oriented programming languages, and integrating with general-purpose programming languages to increase mainstream adoption.
By advancing this paradigm, we can make fixed-point computations more accessible, robust, and efficient for both researchers and practitioners.

\begin{acks}
This work is supported by
the \grantsponsor{singapore-moe}{Singapore Ministry of Education}{https://www.moe.gov.sg/}
under Project Number~\grantnum{singapore-moe}{T1~251RES2422}.
\end{acks}

\bibliographystyle{ACM-Reference-Format}
\bibliography{fixed-point-oriented-programming}


\begin{thebibliography}{23}


\ifx \showCODEN    \undefined \def \showCODEN     #1{\unskip}     \fi
\ifx \showDOI      \undefined \def \showDOI       #1{#1}\fi
\ifx \showISBNx    \undefined \def \showISBNx     #1{\unskip}     \fi
\ifx \showISBNxiii \undefined \def \showISBNxiii  #1{\unskip}     \fi
\ifx \showISSN     \undefined \def \showISSN      #1{\unskip}     \fi
\ifx \showLCCN     \undefined \def \showLCCN      #1{\unskip}     \fi
\ifx \shownote     \undefined \def \shownote      #1{#1}          \fi
\ifx \showarticletitle \undefined \def \showarticletitle #1{#1}   \fi
\ifx \showURL      \undefined \def \showURL       {\relax}        \fi
\providecommand\bibfield[2]{#2}
\providecommand\bibinfo[2]{#2}
\providecommand\natexlab[1]{#1}
\providecommand\showeprint[2][]{arXiv:#2}

\bibitem[Arntzenius and Krishnaswami(2016)]%
        {datafun}
\bibfield{author}{\bibinfo{person}{Michael Arntzenius} {and}
  \bibinfo{person}{Neelakantan~R. Krishnaswami}.}
  \bibinfo{year}{2016}\natexlab{}.
\newblock \showarticletitle{Datafun: a functional Datalog}.
\newblock \bibinfo{journal}{\emph{SIGPLAN Not.}} \bibinfo{volume}{51},
  \bibinfo{number}{9} (\bibinfo{date}{Sept.} \bibinfo{year}{2016}),
  \bibinfo{pages}{214–227}.
\newblock
\showISSN{0362-1340}
\urldef\tempurl%
\url{https://doi.org/10.1145/3022670.2951948}
\showDOI{\tempurl}


\bibitem[Bellman(1958)]%
        {bellman}
\bibfield{author}{\bibinfo{person}{Richard Bellman}.}
  \bibinfo{year}{1958}\natexlab{}.
\newblock \showarticletitle{On a routing problem}.
\newblock \bibinfo{journal}{\emph{Quart. Appl. Math.}}  \bibinfo{volume}{16}
  (\bibinfo{year}{1958}), \bibinfo{pages}{87--90}.
\newblock


\bibitem[Blanchette and Adams(2023)]%
        {fpoponline}
\bibfield{author}{\bibinfo{person}{Henry Blanchette} {and}
  \bibinfo{person}{Michael~D. Adams}.} \bibinfo{year}{2023}\natexlab{}.
\newblock \bibinfo{booktitle}{\emph{Fixed-Point-Oriented Programming
  Language}}.
\newblock
\urldef\tempurl%
\url{https://github.com/rybla/fixlat-lang}
\showURL{%
\tempurl}


\bibitem[Chomsky(1959)]%
        {chomsky}
\bibfield{author}{\bibinfo{person}{Noam Chomsky}.}
  \bibinfo{year}{1959}\natexlab{}.
\newblock \showarticletitle{On certain formal properties of grammars}.
\newblock \bibinfo{journal}{\emph{Information and Control}}
  (\bibinfo{year}{1959}), \bibinfo{pages}{137--167}.
\newblock


\bibitem[Cocke and Schwarts(1970)]%
        {cocke}
\bibfield{author}{\bibinfo{person}{John Cocke} {and} \bibinfo{person}{Jacob~T.
  Schwarts}.} \bibinfo{year}{1970}\natexlab{}.
\newblock \bibinfo{booktitle}{\emph{Programming languages and their compilers:
  Preliminary notes}}.
\newblock \bibinfo{type}{{T}echnical {R}eport}. \bibinfo{institution}{Courant
  Institute of Mathematical Sciences, New York University}.
\newblock


\bibitem[Dijkstra(1959)]%
        {dijkstra}
\bibfield{author}{\bibinfo{person}{Edsger~Wybe Dijkstra}.}
  \bibinfo{year}{1959}\natexlab{}.
\newblock \showarticletitle{A note on two problems in connexion with graphs}.
\newblock \bibinfo{journal}{\emph{Numer. Math.}} (\bibinfo{year}{1959}),
  \bibinfo{pages}{269--271}.
\newblock


\bibitem[Earley(1970)]%
        {10.1145/362007.362035}
\bibfield{author}{\bibinfo{person}{Jay Earley}.}
  \bibinfo{year}{1970}\natexlab{}.
\newblock \showarticletitle{An efficient context-free parsing algorithm}.
\newblock \bibinfo{journal}{\emph{Commun. ACM}} \bibinfo{volume}{13},
  \bibinfo{number}{2} (\bibinfo{date}{Feb.} \bibinfo{year}{1970}),
  \bibinfo{pages}{94–102}.
\newblock
\showISSN{0001-0782}
\urldef\tempurl%
\url{https://doi.org/10.1145/362007.362035}
\showDOI{\tempurl}


\bibitem[Ford(1956)]%
        {ford1}
\bibfield{author}{\bibinfo{person}{L.~R. Ford}.}
  \bibinfo{year}{1956}\natexlab{}.
\newblock \bibinfo{booktitle}{\emph{{Network Flow Theory}}}.
\newblock \bibinfo{type}{{T}echnical {R}eport}. \bibinfo{institution}{RAND
  Corporation}.
\newblock


\bibitem[Ford and Fulkerson(1956)]%
        {ford2}
\bibfield{author}{\bibinfo{person}{L.~R. Ford} {and} \bibinfo{person}{D.~R.
  Fulkerson}.} \bibinfo{year}{1956}\natexlab{}.
\newblock \showarticletitle{{Maximal Flow Through a Network}}.
\newblock \bibinfo{journal}{\emph{Canadian Journal of Mathematics}}
  \bibinfo{volume}{8} (\bibinfo{year}{1956}), \bibinfo{pages}{399--404}.
\newblock


\bibitem[Gallaire and Minker(1977)]%
        {datalog}
\bibfield{author}{\bibinfo{person}{Hervé Gallaire} {and} \bibinfo{person}{Jack
  Minker}.} \bibinfo{year}{1977}\natexlab{}.
\newblock \showarticletitle{{Logic and Data Bases}}.
\newblock \bibinfo{journal}{\emph{Symposium on Logic and Data Bases, Centre
  d'études et de recherches de Toulouse, France, 1977}}
  (\bibinfo{year}{1977}).
\newblock


\bibitem[Hopcroft(1971)]%
        {10.5555/891883}
\bibfield{author}{\bibinfo{person}{John~E. Hopcroft}.}
  \bibinfo{year}{1971}\natexlab{}.
\newblock \bibinfo{booktitle}{\emph{An n log n algorithm for minimizing states
  in a finite automaton}}.
\newblock \bibinfo{type}{{T}echnical {R}eport}. \bibinfo{address}{Stanford, CA,
  USA}.
\newblock


\bibitem[Jaffar and Maher(1994)]%
        {DBLP:journals/jlp/JaffarM94}
\bibfield{author}{\bibinfo{person}{Joxan Jaffar} {and}
  \bibinfo{person}{Michael~J. Maher}.} \bibinfo{year}{1994}\natexlab{}.
\newblock \showarticletitle{Constraint Logic Programming: {A} Survey}.
\newblock \bibinfo{journal}{\emph{The Journal of Logic Programming}}
  \bibinfo{volume}{19/20} (\bibinfo{year}{1994}), \bibinfo{pages}{503--581}.
\newblock
\showISSN{0743-1066}
\urldef\tempurl%
\url{https://doi.org/10.1016/0743-1066(94)90033-7}
\showDOI{\tempurl}
\newblock
\shownote{Special Issue: Ten Years of Logic Programming}.


\bibitem[Jordan et~al\mbox{.}(2016)]%
        {DBLP:conf/cav/JordanSS16}
\bibfield{author}{\bibinfo{person}{Herbert Jordan}, \bibinfo{person}{Bernhard
  Scholz}, {and} \bibinfo{person}{Pavle Suboti{\'{c}}}.}
  \bibinfo{year}{2016}\natexlab{}.
\newblock \showarticletitle{Souffl{\'{e}}: On Synthesis of Program Analyzers}.
  In \bibinfo{booktitle}{\emph{Computer Aided Verification}}
  \emph{(\bibinfo{series}{Lecture Notes in Computer Science},
  Vol.~\bibinfo{volume}{9780})}, \bibfield{editor}{\bibinfo{person}{Swarat
  Chaudhuri} {and} \bibinfo{person}{Azadeh Farzan}} (Eds.).
  \bibinfo{publisher}{Springer}, \bibinfo{pages}{422--430}.
\newblock
\showISBNx{978-3-319-41540-6}
\urldef\tempurl%
\url{https://doi.org/10.1007/978-3-319-41540-6\_23}
\showDOI{\tempurl}


\bibitem[Kasami(1965)]%
        {kasami}
\bibfield{author}{\bibinfo{person}{Tadao Kasami}.}
  \bibinfo{year}{1965}\natexlab{}.
\newblock \bibinfo{booktitle}{\emph{An efficient regonition and syntax-analysis
  algorithm for context-free languages}}.
\newblock \bibinfo{type}{{T}echnical {R}eport}. \bibinfo{institution}{AFCRL}.
\newblock


\bibitem[Knuth and Plass(1981)]%
        {https://doi.org/10.1002/spe.4380111102}
\bibfield{author}{\bibinfo{person}{Donald~E. Knuth} {and}
  \bibinfo{person}{Michael~F. Plass}.} \bibinfo{year}{1981}\natexlab{}.
\newblock \showarticletitle{Breaking paragraphs into lines}.
\newblock \bibinfo{journal}{\emph{Software: Practice and Experience}}
  \bibinfo{volume}{11}, \bibinfo{number}{11} (\bibinfo{year}{1981}),
  \bibinfo{pages}{1119--1184}.
\newblock
\urldef\tempurl%
\url{https://doi.org/10.1002/spe.4380111102}
\showDOI{\tempurl}
\showeprint{https://onlinelibrary.wiley.com/doi/pdf/10.1002/spe.4380111102}


\bibitem[Lengauer and Tarjan(1979)]%
        {10.1145/357062.357071}
\bibfield{author}{\bibinfo{person}{Thomas Lengauer} {and}
  \bibinfo{person}{Robert~Endre Tarjan}.} \bibinfo{year}{1979}\natexlab{}.
\newblock \showarticletitle{A fast algorithm for finding dominators in a
  flowgraph}.
\newblock \bibinfo{journal}{\emph{ACM Trans. Program. Lang. Syst.}}
  \bibinfo{volume}{1}, \bibinfo{number}{1} (\bibinfo{date}{Jan.}
  \bibinfo{year}{1979}), \bibinfo{pages}{121–141}.
\newblock
\showISSN{0164-0925}
\urldef\tempurl%
\url{https://doi.org/10.1145/357062.357071}
\showDOI{\tempurl}


\bibitem[Madsen et~al\mbox{.}(2016)]%
        {flix}
\bibfield{author}{\bibinfo{person}{Magnus Madsen}, \bibinfo{person}{Ming-Ho
  Yee}, {and} \bibinfo{person}{Ond\v{r}ej Lhot\'{a}k}.}
  \bibinfo{year}{2016}\natexlab{}.
\newblock \showarticletitle{From Datalog to flix: a declarative language for
  fixed points on lattices}. In \bibinfo{booktitle}{\emph{Proceedings of the
  37th ACM SIGPLAN Conference on Programming Language Design and
  Implementation}} (Santa Barbara, CA, USA) \emph{(\bibinfo{series}{PLDI
  '16})}. \bibinfo{publisher}{Association for Computing Machinery},
  \bibinfo{address}{New York, NY, USA}, \bibinfo{pages}{194–208}.
\newblock
\showISBNx{9781450342612}
\urldef\tempurl%
\url{https://doi.org/10.1145/2908080.2908096}
\showDOI{\tempurl}


\bibitem[Okhotin(2004)]%
        {booleangrammars}
\bibfield{author}{\bibinfo{person}{Alexander Okhotin}.}
  \bibinfo{year}{2004}\natexlab{}.
\newblock \showarticletitle{Boolean grammars}.
\newblock \bibinfo{journal}{\emph{Information and Computation}}
  (\bibinfo{year}{2004}), \bibinfo{pages}{19--48}.
\newblock


\bibitem[Sakai(1961)]%
        {syntaxinuniversaltranslation}
\bibfield{author}{\bibinfo{person}{Itiroo Sakai}.}
  \bibinfo{year}{1961}\natexlab{}.
\newblock \showarticletitle{Syntax in universal translation}. In
  \bibinfo{booktitle}{\emph{International Conference on Machine Translation of
  Languages and Applied Language Analysis}}.
\newblock


\bibitem[Scholz et~al\mbox{.}(2016)]%
        {DBLP:conf/cc/ScholzJSW16}
\bibfield{author}{\bibinfo{person}{Bernhard Scholz}, \bibinfo{person}{Herbert
  Jordan}, \bibinfo{person}{Pavle Suboti\'{c}}, {and} \bibinfo{person}{Till
  Westmann}.} \bibinfo{year}{2016}\natexlab{}.
\newblock \showarticletitle{On fast large-scale program analysis in Datalog}.
  In \bibinfo{booktitle}{\emph{Proceedings of the 25th International Conference
  on Compiler Construction}} (Barcelona, Spain) \emph{(\bibinfo{series}{CC
  '16})}, \bibfield{editor}{\bibinfo{person}{Ayal Zaks} {and}
  \bibinfo{person}{Manuel~V. Hermenegildo}} (Eds.).
  \bibinfo{publisher}{Association for Computing Machinery},
  \bibinfo{address}{New York, NY, USA}, \bibinfo{pages}{196--206}.
\newblock
\showISBNx{978-1-4503-4241-4}
\urldef\tempurl%
\url{https://doi.org/10.1145/2892208.2892226}
\showDOI{\tempurl}


\bibitem[Sheard and Jones(2002)]%
        {10.1145/581690.581691}
\bibfield{author}{\bibinfo{person}{Tim Sheard} {and}
  \bibinfo{person}{Simon~Peyton Jones}.} \bibinfo{year}{2002}\natexlab{}.
\newblock \showarticletitle{Template meta-programming for Haskell}. In
  \bibinfo{booktitle}{\emph{Proceedings of the 2002 ACM SIGPLAN Workshop on
  Haskell}} (Pittsburgh, Pennsylvania) \emph{(\bibinfo{series}{Haskell '02})}.
  \bibinfo{publisher}{Association for Computing Machinery},
  \bibinfo{address}{New York, NY, USA}, \bibinfo{pages}{1–16}.
\newblock
\showISBNx{1581136056}
\urldef\tempurl%
\url{https://doi.org/10.1145/581690.581691}
\showDOI{\tempurl}


\bibitem[Tarjan(1972)]%
        {robert1972}
\bibfield{author}{\bibinfo{person}{Robert Tarjan}.}
  \bibinfo{year}{1972}\natexlab{}.
\newblock \showarticletitle{{Depth-First Search and Linear Graph Algorithms}}.
\newblock \bibinfo{journal}{\emph{SIAM J. Comput.}} \bibinfo{volume}{1},
  \bibinfo{number}{2} (\bibinfo{year}{1972}), \bibinfo{pages}{146--160}.
\newblock


\bibitem[Younger(1967)]%
        {younger}
\bibfield{author}{\bibinfo{person}{Daniel~H. Younger}.}
  \bibinfo{year}{1967}\natexlab{}.
\newblock \showarticletitle{Recognition and parsing of context-free languages
  in time $n^3$}.
\newblock \bibinfo{journal}{\emph{Information and Control}}
  \bibinfo{volume}{10}, \bibinfo{number}{2} (\bibinfo{year}{1967}),
  \bibinfo{pages}{189--208}.
\newblock


\end{thebibliography}

\end{document}